# *Ab initio* random structure searching for battery cathode materials

Ziheng Lu[1,*,†], Bonan Zhu[2,*,†], Benjamin W. B. Shires[1], David O. Scanlon[2,3,4], Chris J. Pickard[1,5]

**AFFILIATIONS**

[1] Department of Materials Science & Metallurgy, University of Cambridge, 27 Charles Babbage Road, Cambridge, CB3 0FS, UK

[2] Department of Chemistry, University College London, 20 Gordon Street, London WC1H 0AJ, UK

[3] Thomas Young Centre, University College London, Gower Street, London WC1E 6BT, UK

[4] Diamond Light Source Ltd., Diamond House, Harwell Science and Innovation Campus, Didcot, Oxfordshire OX11 0DE, UK

[5] Advanced Institute for Materials Research, Tohoku University, Sendai, Japan

[†] These authors contributed equally to this work

[*]Author to whom correspondence should be addressed: zl462@cam.ac.uk bonan.zhu@ucl.ac.uk

**ABSTRACT**

Cathodes are critical components of rechargeable batteries. Conventionally, the search for cathode materials relies on experimental trial-and-error and a traversing of existing computational/experimental databases. While these methods have led to the discovery of several commercially-viable cathode materials, the chemical space explored so far is limited and many phases will have been overlooked, in particular those that are metastable. We describe a computational framework for battery cathode exploration, based on *ab initio* random structure searching (AIRSS), an approach that samples local minima on the potential energy surface to identify new crystal structures. We show that, by delimiting the search space using a number of constraints, including chemically aware minimum interatomic separations, cell volumes, and space group symmetries, AIRSS can efficiently predict both thermodynamically stable and metastable cathode materials. Specifically, we investigate $LiCoO_2$, $LiFePO_4$, and $Li_xCu_yF_z$ to demonstrate the efficiency of the method by rediscovering the known crystal structures of these cathode materials. The effect of parameters, such as minimum separations and symmetries, on the efficiency of the sampling is discussed in detail. The adaptation of the minimum interatomic distances, on a species-pair basis, from low-energy optimized structures to efficiently capture the local coordination environment of atoms, is explored. A family of novel cathode materials based, on the transition-metal oxalates, is proposed. They demonstrate superb energy density, oxygen-redox stability, and lithium diffusion properties. This article serves both as an introduction to the computational framework, and as a guide to battery cathode material discovery using AIRSS.

## I. INTRODUCTION

Lithium-ion batteries (LIBs) are electrochemical energy storage devices characterized by high energy density, good rate capability, and a long shelf-life. They have dominated the market for portable electronics since their commercialization in the 90s.[1,2] This technology has been challenged by the increasing demands for higher energy density, due to the growing electric vehicle market, and grid-scale energy storage. One of the most important performance parameters, the energy density of a battery is determined by its average discharge voltage, lithium storage capacity, and weight (or volume). Among these factors, both the voltage and the capacity depend heavily on the physiochemical properties of the cathode materials used. Therefore, the search for cathode materials with high discharge voltages and large capacities is one of the central aims of battery research.[3,4] Beyond that, due to the demand for high-power output by electric vehicles and grid-scale energy storage, a high-rate capability of a battery is also essential. In this context, the cathodes that transport both lithium and electrons rapidly are critically needed.

Conventionally, new cathode materials are discovered through *ad hoc* design, based on the understanding of known chemistry and crystallography. For example, the intercalation reaction between guest ions with solid hosts has been known for over a century. [4] By studying the fundamental properties of the $Li_xTiS_2$ series, Whittingham used the layered $TiS_2$ as a cathode, and demonstrated the first rechargeable lithium battery. [5] However, such cathodes suffer from low redox potential versus $Li/Li^+$ so the energy density must be improved. To achieve this, Goodenough and co-workers replaced the $S^{2-}$ with $O^{2-}$ and enabled the use of the $Co^{3+/4+}$ redox couple, based on the knowledge that the $S^{2-}$ 3p band lies at a higher energy as compared with that of $O^{2-}$ 2p. [6] This led directly to the discovery of $LiCoO_2$, one of the most successful cathode materials for LIBs to date. [7] Despite the commercial success, the search for cathode materials based on *ad hoc* design and experimental trial-and-error has proven difficult. In fact, only a very limited number of materials have been confirmed to be competitive, despite the tremendous efforts made in searching for them. A new paradigm for cathode material discovery is critically needed.

Computational materials science has become a powerful tool in materials discovery. [8-10] Thanks to developments in electronic structure methods, in particular density functional theory (DFT) and related computational approaches, the first-principles computation of the properties of practical materials have become routine in materials research. [11-14] At the same time, the dramatic increase in computational capacity of our research infrastructure has empowered researchers to carry out high-throughput screening of materials for those with desired properties. [8] Databases have been assembled to archive properties such as formation energies and band gaps. [15, 16] In the context of cathode materials, phase stability, discharge voltage, and cation diffusion properties can be predicted and analyzed based on these precomputed data entries. [17, 18] While such an approach has led to the discovery of a number of cathodes, [19-21] these data records typically derive from experimental reports and are limited in number. Therefore, the compositional and structural space explored is severely constrained, and many possible combinations of chemical species have been overlooked. More importantly, even for a single given composition, crucial information is omitted that would aid in the exploration of new materials – namely polymorphism, or different low energy structures. These metastable materials can sometimes be synthesized and display desirable properties. In fact, materials with polymorphs have been widely studied in a number of areas including high pressure physics, organic chemistry, and electrochemistry. [22-24] In the context of battery cathode materials, such polymorphism may provide exciting opportunities for materials with extraordinary electron and ion transport characteristics.

Here we address these issues by showing how *ab initio* random structure searching (AIRSS) can be used as an efficient tool for the exploration of novel cathode materials for batteries. [25-27] The method is based on a random sampling of the first principles potential energy surfaces (PESs) of atomic arrangements in solids and has been successfully applied in a number of areas including high pressure physics, superconductors, and semiconductors. [26] Here, we demonstrate that by constraining the search space through the choice of parameters such as chemically aware minimum interatomic distances, cell volumes, and symmetries, the efficient identification of both stable and metastable cathode materials is possible. Specifically, we explore existing cathode compositions such as $LiCoO_2$, $LiFePO_4$, and $Li_xCu_yF_z$ to showcase this method and show how to achieve an efficient sampling of the PES. Building on this, we apply the AIRSS method to lithium-stuffed transition-metal oxalates in the search for novel cathode materials. Screening the low energy outcomes of the search, we identify several oxalate polymorphs as good candidates, with decent rate capability and energy density.

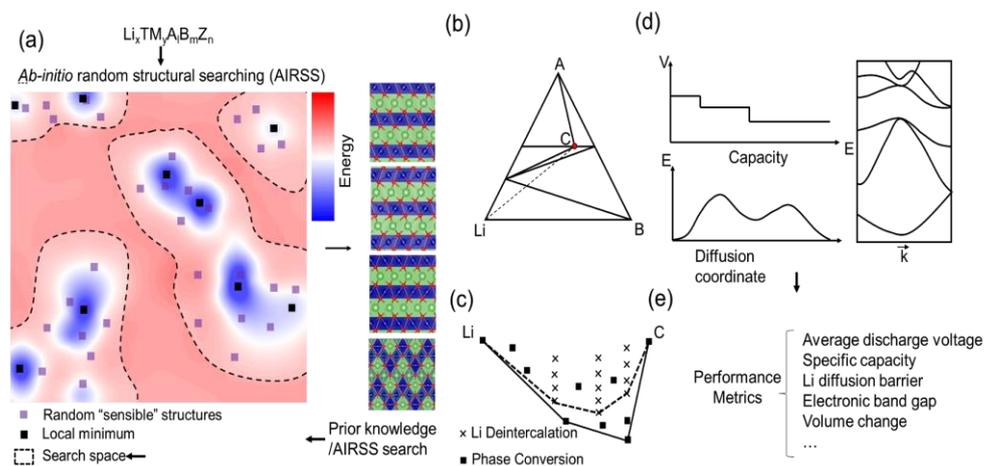

**Fig. 1** Schematic of the AIRSS-based framework for the discovery of new battery cathodes: (a) Sampling the potential energy surface with AIRSS, construction of (b) the compositional phase diagram and (c) the pseudo-binary convex hull along the Li-cathode tie line, (d) post-search computation of cathode properties that can be derived from DFT calculations (e.g. voltages, ion diffusion barriers, electronic band gaps), and (e) evaluation of performance metrics for battery cathodes.

## II. METHODOLOGY

### 1. General searching framework

The general searching framework is illustrated in **Fig 1**. It consists of the choice of composition, the crystal structure search using AIRSS, and the post-search screening. Composition-wise, we chose $LiCoO_2$, $LiFePO_4$, $Li_xCu_yF_z$, and $Li_2TM(C_2O_4)_2$ (TM=Fe, Co, Ni, V, and Mn) in this study. $LiCoO_2$ and $LiFePO_4$ were selected because they are the most representative materials of two cathode families: simple transition metal oxides and polyanionic compounds. It is worth noting that, to determine the stability of a phase, not only the composition of interest needs to be considered, but also all the other possible ones within the given chemical space. While, for most systems, the stable phases can be directly taken from databases, for systems that are less studied, an extensive search over the entire compositional space is necessary to construct the relevant phase diagram. In this case, we demonstrate the search of a phase diagram using $Li_xCu_yF_z$ because $Cu_2F$ is a high-capacity conversion-type cathode, which undergoes interesting conversion reactions during charge and discharge. Finally, $Li_2TM(C_2O_4)_2$ was chosen to highlight the capability of the current searching scheme because $Li_2Fe(C_2O_4)$ has been reported to be a potential cathode with interesting anion redox properties while all the other members are less well studied.

After choosing the composition, crystal structure predictions were carried out using AIRSS. The details of this approach are provided in the next subsection, and the impact of selecting the few search parameters on its efficiency is discussed in section 3.1. After the AIRSS search, candidate structures were selected for property screening based on their phase stability. For a cathode material, the properties of interest are the voltage, the capacity, the cation transport barrier, and the electronic band gaps. The first two determine the energy density of a cathode while the latter two provide a proxy to predict the rate capability. These property calculations are standard in battery research and are briefly discussed in section 2.3.

**2. *Ab initio* random structure searching**

At the core of the searching scheme, AIRSS samples the local energy minima of a PES by first generating random "sensible" structures, and then locally optimising their geometries using first-principles calculations. The efficiency or even the success of the algorithm, as we will explain later, is highly dependent on choosing sensible parameters and constraints based on physically quantities, *e.g.,* reasonable density, no close contacts, and proper neighbor relations. The general workflow of an AIRSS search is depicted in **Fig 1(a)**. Generating random structures with few constraints will provide the widest coverage of the search space and therefore reduces the possibility of completely missing out the global minimum. However, it will also result in low search efficiency, i.e. the rate of encountering lower energy structures is reduced. Considering that cathode materials operate at ambient pressures, it is natural to use chemical ideas to constrain the search space and steer the search towards chemically "sensible" regions. This is done by applying constraints during structure generation. To generate a structure that is chemically sensible, both the bond lengths and the neighboring relation between atoms need to be constrained. Two parameters are effective in this context. One is the minimum separation between atoms (which may be different for each pair of species, and so chemically aware). To acquire structures that fulfil the distance constraints, they are generated randomly and are optimized using a hard sphere potential to reach the desired species-dependent minimum separations. It is important to note that such minimum separations are species-pairwise and an example is shown in **Table 1**. Such distance constraints avoid close contacts between atoms which may lead to numerical problems for the first principles code during the geometry relaxation. Furthermore, by fixing the cell volume (per formula unit/f.u.), such species-pairwise minimum separations can be used to avoid unwanted neighboring relations. For example, by setting the minimum separations between Li and Li and between Li and O to 2.8Å and 2.0Å, respectively in a lithium containing oxide, $OLi_x$ clusters will be favored instead of $Li_x$ neighboring. Should one have knowledge on the chemical system, the values of the minimum separations and cell volumes can be measured from known structures. For the case where the chemistry is new, one can perform a preliminary AIRSS search using random minimum separations, and then measure the resulting values from the structure with the lowest energy. In practice, to avoid overly constraining the system, the structural rejection criteria based on the minimum separation matrix are slightly loosened. Symmetry is also exploited during the AIRSS search as most known crystals are characterized by a degree of symmetry. Moreover, adopting symmetry accelerates the DFT calculations through reducing the number of k-points required to sample the Brillouin Zone, and the number of degrees of freedom in the local geometry optimisation. In an AIRSS search, one can generate initial structures with a specified number of randomly chosen symmetry operations. The effect of symmetry will be discussed further in section 3.1.1.

As a random sampling approach, there is no rigorous convergence criteria to tell when to stop an AIRSS search. However, it is clear when the search should not be stopped. Searching should continue at least until the following requirements are fulfilled: 1) known marker structures, if any, are identified; 2) a number of structures with low energies are encountered; and 3) these structures are encountered multiple times.

## 3. Density functional theory calculations

As part of an AIRSS search, the geometry optimization and the energy evaluations are carried out based on first principles DFT calculations. In this work, the plane-wave DFT code CASTEP was used for each geometry optimization. [28, 29] The Perdew–Burke–Ernzerho (PBE) exchange-correlation functionals and the on-the-fly generated ultrasoft pseudopotentials (QC5) were adopted. [29, 30] A relatively low plane-wave cut-off energy of 340 eV was used. The reciprocal space was sampled on Monkhorst-Pack grids with a spacing of 0.07 Å$^{-1}$. The effect of spin-polarization and the Hubbard U correction on the sampling of the PES will be discussed in section 3.1.1.

For further structural optimization and property calculations, spin-polarized DFT

calculations were performed using PBE exchange-correlation functionals with Hubbard U corrections. For this we used the VASP code to make the use of the calibrated U parameters for phase stability and voltage prediction. [31-35] A plane-wave cut-off energy of 520 eV was used, alongside Monkhorst-Pack grids with a spacing of 0.04 Å$^{-1}$. Hubbard U corrections of 5.3 eV, 6.2 eV, 3.25 eV, 3.9 eV, and 3.32eV, were applied to the d-channel of Fe, Ni, V, Mn, Co, respectively. [33, 35] The band gap values were estimated using the HSE06 hybrid functional based on the PBE+U relaxed structures. [36] Ferromagnetic spin arrangements are assumed in spin-polarized DFT calculations.

## 4. Property calculations

The average cathode voltage $V_a$ is calculated using the following equation:

$$V_a = \frac{-[E_{Cat-Li_{x_j}} - E_{Cat-Li_{x_i}} - (x_j - x_i)E_{Li}]}{(x_j - x_i)e}$$

where $E_{Cat-Li_x}$ and $E_{Li}$ are the enthalpy per atom of the cathode with a Li concentration of $x$, and the enthalpy per atom of lithium metal, respectively. $e$ is the charge of an electron.

The capacity of a cathode is relatively difficult to accurately determine from simple DFT calculations, because this quantity is dependent on the experimental charge and discharge processes, as well as the relative stability of the cathode during such a process. We can estimate an upper bound by looking at the stability of the relaxed structures of the lithiated phases by removing lithium from the structures followed by structural relaxations. The gravimetric energy density of the cathode is estimated by multiplying the average voltage and the capacity followed by normalization by its weight.

The rate capability of cathodes can be estimated by calculating the Li diffusion barrier and the band gap, which give information about the ion and the electron transport, respectively. The Li diffusion barrier was estimated using climbing image nudged elastic band calculations, [37] and the band gaps were calculated using the HSE06 functional as described previously. [36]

**5. PES visualization**

To present the results of our searches, we generated structure maps depicting the relative positions of each structure on the underlying PES. To do this, we first generated the Smooth Overlap of Atomic Positions (SOAP) [38][39] description of each structure in a given data-set using the ASAP code, [40] characterizing each structure as a vector in a high-dimensional space. We used the universal SOAP parameters in ASAP, which are constructed based on the elements present in the structures, following the heuristics of Cheng *et al.* [40] Note that we included cross-over terms between elements in these descriptors, and the global description of each structure is yielded by averaging over the local environments on an element-wise basis. The descriptor vectors were normalised so that each variable had a variance of unity across a given data-set.

We then applied the dimensionality reduction method Stochastic Hyperspace Embedding And Projection (SHEAP) to the high-dimensional data, to produce two-dimensional representations. [41,42] SHEAP constructs a weighted graph describing the similarity relationships in the source data. Following a test to combine equivalent structures, the data is cast into a low-dimensional space (2D in this case) using random projection. Another graph of weights is then constructed for the nodes defined by these mapped points. The layout of the map is then optimized by minimizing a cost-function which penalizes any mismatch between the weights.

In these maps, individual structures are represented by circles colored according to enthalpy, with areas proportional to the number of occurrences in the search. Circles are prevented from overlapping each other by a soft-core repulsion that is turned on only once the nearly optimal layout has been reached. The axes do not have any predetermined physical interpretation, but the relative positioning and clustering across the two-dimensional space reveals the structural similarity.

## III. RESULTS AND DISCUSSION

### 1. Re-exploring known systems and the effect of searching parameters

### 1.1 LiCoO$_2$ - impact of the species-pairwise minimum separations, cell volumes, and DFT parameters on search efficiency

We start by performing an AIRSS search on LiCoO$_2$, the most commonly used cathode material in LIBs. For this system, we conducted the search using the minimum separation matrix and volume per formula measured from the known $R\bar{3}m$ high-temperature (HT) phase.[43, 44] The values are listed in **Table 1**. The number of formula units was chosen to take values of 1 to 5. The results of this search are summarized in **Fig. 2(a)**. In total, 1200 random structures were generated and optimized. Of the relaxed structures, the trigonal-layered HT phase had the highest encounter rate, at 47%. The cubic-lithiated-spinel low-temperature (LT) phase, which has a comparable energy, was also found multiple times (~2%).[45] Other low energy structures found include the P2 and O2 phases of LiCoO$_2$, which result from slight variation of the CoO$_6$ layer stacking, compared to the O3-type stacking of HT-LiCoO$_2$.[46, 47] All of these low energy structures are characterized by similar local coordination - cobalt ions are octahedral, whilst lithium ions are either octahedral or prismatic. This search also turned up many polymorphs with relatively high energies. These include several approximants of disordered rocksalts, in which cobalt and lithium ions are distributed in a non-layered manner, within a distorted MO$_6$ (M=Li and Co) framework. In these structures, the large lattice distortion gives rise to the higher energies. We expect these disordered rocksalts to be difficult to synthesize.

**Table 1** Species-pairwise minimum atomic separations and cell volume of the $R\bar{3}m$ high-temperature phase of LiCoO$_2$.

| Species pair | Minimum separation Å | Cell volume Å$^3$ fu$^{-1}$ |
|:---:|:---:|:---:|
| Li-Li | 2.81 | |
| Li-O | 2.04 | |
| Li-Co | 2.80 | 25.8 |
| O-O | 2.62 | |
| O-Co | 1.93 | |
| Co-Co | 2.82 | |

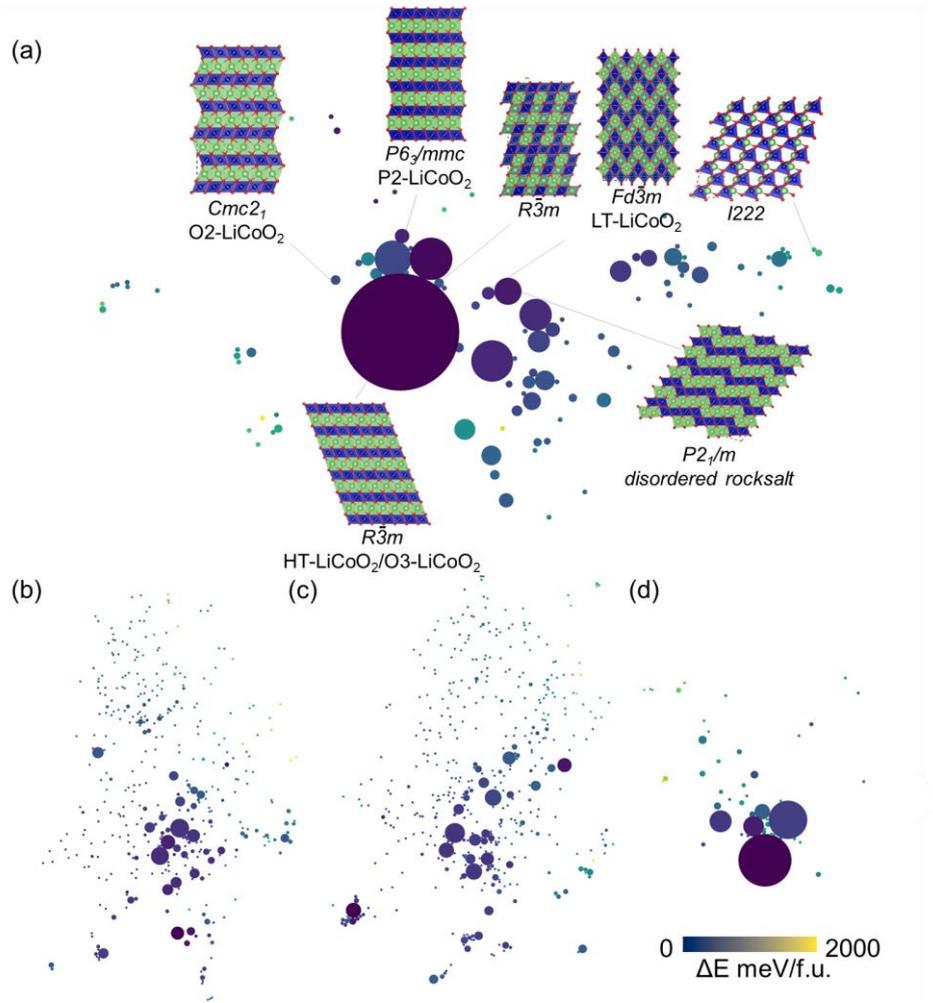

**Fig. 2** SHEAP map of the local energy minima of LiCoO$_2$ using different search parameters: (a) with the minimum separation matrix derived from the experimental structure, imposing 2-4 symmetry operations, (b) with random minimum separations from 1Å to 3Å, imposing 2-4 symmetry operations, (c) with fixed minimum separations of 1.5Å, imposing 2-4 symmetry operations and (d) with minimum separations derived from the experimental structure, imposing no symmetry constraints.

The above-mentioned search was carried out based on prior knowledge of LiCoO$_2$. However, in exploratory searches, such prior knowledge may not exist. Now, we demonstrate that constraining the search space using the minimum separation matrix, cell volumes, and symmetry is critical to achieving efficient sampling. This is followed by a discussion on how one can use a preliminary AIRSS search to determine a suitable list of minimum separation values and cell volumes.

**Figs. 2(b)** and **3(c)** summarize search results obtained from the use of a matrix of minimum separations drawn uniformly from 1Å-3Å, and of a single fixed minimum separation for all species pairs of 1.5Å, respectively. In comparison to **Fig. 2(a)**, these searches resulted in many more high energy configurations, with a significantly lower encounter rate for the lowest energy structures. In the case of a single fixed minimum separation we can easily understand this by noting that the chemical environments of the cation and anion are not distinguished by this constraint. As a result, Li-Li and O-O close contacts are generated resulting in unphysical and high energy structures and chemistries.

As illustrated by the results of **Fig. 2**, an effective way to select suitable minimum separation values is to measure them from known structures with similar chemical compositions. However, in some cases, the detailed coordination environment is not known as a prior and the minimum separations need to be selected using a first principles approach. A way around this is through a preliminary AIRSS search. **Fig. 3** shows the distribution of minimum separations arising in relaxed structures obtained from a search on $LiCoO_2$ using a random minimum separation matrix whose elements are drawn uniformly from 1-3Å. No symmetry constraints were applied. We observed, for each pair of species, rapid convergence in the lower bound of the interatomic distances with respect to increasing the number of formula units, with the converged values corresponding to the minimum bond lengths of the system. Thus, for a "safe" search, one can use a relatively small cell to acquire these minimum separations, ruling out certain close contacts between pairs of species. However, these lower bounds do not consider the detailed local coordination of the system. For example, as shown in the inset of **Fig. 3**, the lower bound of Co-Co distance is ~2.1 Å, which corresponds to Co pairs in an O-connected $CoO_4$-$CoO_4$ configuration. Compared with the $CoO_6$ octahedra, such local coordination is energetically less favorable. Therefore, for an "efficient" search, it is better to choose the minimum separations from the structure with the lowest energy, as indicated by the dashed lines in **Fig. 3**. This will bias the search towards having local coordinations that are more likely to occur in low energy structures.

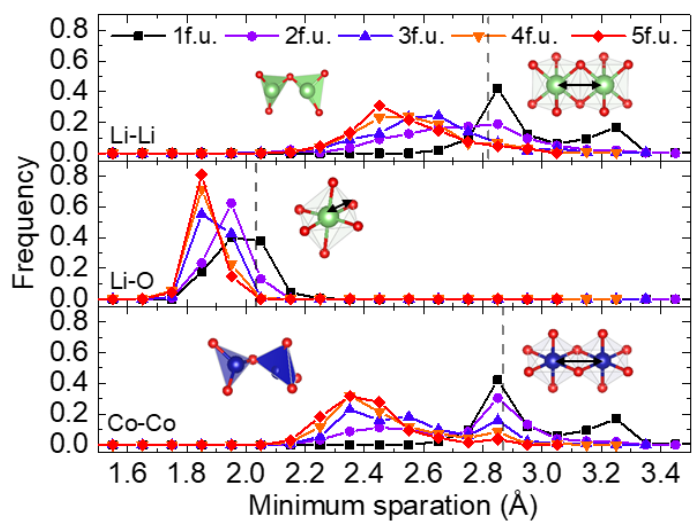

**Fig. 3** The distribution of minimum atomic separations arising in AIRSS searches for $LiCoO_2$, using different numbers of formula units. The dashed lines show the minimum separations in the lowest energy structures.

A search can be further speed up through the application of symmetry constraints. As illustrated in **Figs. 2(a) and 2(d),** for a well-chosen minimum separation matrix, the encounter rates for low energy structures are comparably high with and without the use of symmetry. However, the overall searching time is significantly longer (~10 times) in the absence of symmetry, due to slower DFT electronic steps during the geometry optimizations.

It is worth noting that the majority of the time taken for an AIRSS search is spent on the first principles geometry optimizations. Thus, speeding up the DFT calculations is highly beneficial. One way we achieved this was by adopting ultrasoft pseudopotentials (QC5) specially-made for high-throughput calculations, which allow the use of low energy cutoff. Apart from that, the inclusion of spin polarization and Hubbard U parameters also affects the speed of calculation. Therefore, another way to accelerate the process is to perform non-spin polarized calculations during the search followed by a subsequent round of high accuracy calculations to refine the low energy structures. For $LiCoO_2$, the structural density of states obtained without spin, with spin, and with the Hubbard U correction are similar, and the low energy structures are all found with comparable encounter rates, as shown in **Fig. 4a**. **Fig. 4b** further shows the correspondence of energies obtained with non-spin polarized PBE and with spin-polarized PBE+U for the structures of $LiCoO_2$. We observe a shift of stability order with an energy variation of ~150 meV/atom. [32, 33] For an AIRSS search carried out using non-spin polarized PBE, any structure with energy below this value would need to be recalculated to obtain the correct ordering. The magnitude of this threshold is of course system dependent. In extreme cases, the choice of DFT parameters may significantly alter the PES, possibly resulting in missing certain local minima.

In summary, our workflow for efficient AIRSS is as follows: First, obtain the species-pairwise minimum separations and the cell volume parameters which constrain the search space, either by considering prior knowledge of the chemical system, or by carrying out a preliminary search using random minimum separations; then, generate and relax random sensible structures with different numbers of formula units.

Particularly in the case of many formula units, symmetry may need to be taken into consideration to accelerate the search. Additional re-optimization of low energy structures may need to be carried out if the search used relatively strong assumptions for the DFT calculations.

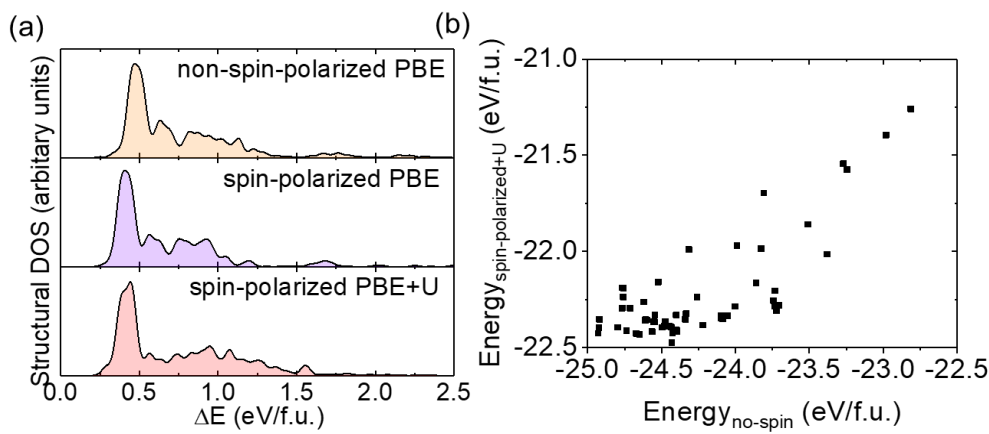

**Fig 4** (a) Structural density of states of AIRSS searches for LiCoO$_2$ using different DFT parameters. (b) The energy correspondence of structures optimized using non-spin polarized PBE and spin-polarized PBE+U.

**1.2 LiFePO$_4$ - effect of adopting structural units on search efficiency**

LiFePO$_4$ is a representative polyanion cathode material that is widely used in electric vehicles due to its low raw material cost. [48] LiFePO$_4$ is structurally more complicated than LiCoO$_2$ and therefore is a more challenging task for crystal structure prediction. In our application of AIRSS to this system, we generated random structures fulfilling the following requirements: 1) the structure has 2 to 4 symmetry operations; 2) the P and O species are generated as proper PO$_4$ units; 3) the interatomic distances are larger than the predefined species-pairwise minimum separation values, which are obtained from the known olivine phase (the P-O distances are specially treated so that only those measured from inter-PO$_4$ units are constrained to fulfil this requirement); 4) the volume per formula unit is within 15% as compared with that of the known olivine phase. The search results, obtained from relaxing these random sensible structures, are summarised in **Fig. 5.** Out of the 684 unique structures found, the experimentally observed olivine phase ($Pnma$) turned up 4 times, giving an encounter rate of ~0.58%. Also found was a $Cmcm$ structure, slightly higher in energy than the olivine phase, that has previously been reported under high pressure conditions.[49] Interestingly, many other phases with similar energies to this were located. Some of these are likely to be synthesizable under fine-tuned conditions, and thus may be worth further computational or experimental study.

In our study of this system, we observed that the use of pre-defined chemical subunits (PO$_4$), as opposed to individual atoms (P and O), was effective in steering the search towards lower energy structures. This is illustrated in the inset of **Fig. 5**, through the structural density of states obtained with and without this constraint. Caution is needed with such constraints, however, since they restrict the sampling space. They should be used only when the chemistry is well understood.[50] In the case of LiFePO$_4$, structures containing other polyphosphate anions are not energetically competitive, and so can be disregarded, allowing us to accelerate the search by imposing P and O to be present as tetrahedral units of PO$_4$.

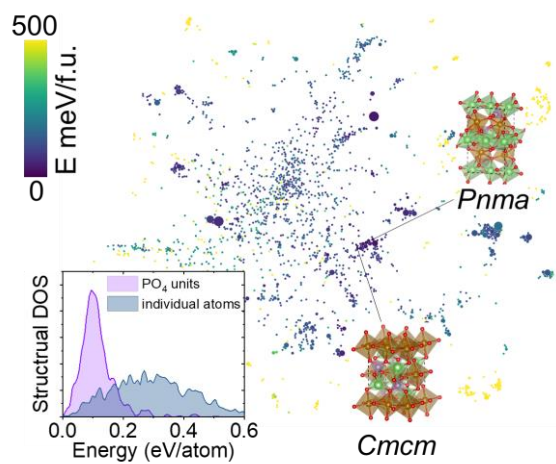

**Fig. 5** SHEAP map of the AIRSS search results for LiFePO$_4$. Inset shows the structural density of states of LiFePO$_4$ generated by treating P and O as PO$_4$ units and as individual atoms.

**1.3 Li$_x$Cu$_y$F$_z$ – a multi-compositional search for phase stability**

The cases of LiCoO$_2$ and LiFePO$_4$ demonstrated the power of AIRSS for compounds of a given single composition. However, in exploratory searches, the determination of the stability of an unknown phase not only involves the calculation of its formation energy, but also whether it is susceptible to decomposition into other phases. Therefore, it is necessary to predict the entire compositional phase diagram. For cathode materials, this is especially important when the chemistry of the system is less understood. The construction of the phase diagram is critical in determining the reaction pathways during lithiation, since not all cathodes are intercalation-type and topotactic phase transformation may happen. For example, transition metal fluorides, for example FeF$_2$, FeF$_3$, and CuF$_2$, are typical cathode materials that go through a conversion reaction during discharge. [51] Here we consider the system of Li-Cu-F, demonstrating how a ternary phase diagram can be generated by conducting an extensive multi-composition search in the chemical space.[52] In particular, we searched for structures with

compositions with the formula $Li_xCu_yF_z$ where x, y, z fulfill the following requirements:

$$x + y + z = n, n < 10$$

and

$$x, y, z \in \mathbb{Z}_+$$

where $\mathbb{Z}_+$ is the set of positive integers. This led to 163 compositions. For each composition, we generated at least 50 random structures for further first principles optimization. In total, 15,031 structures were generated and optimized. This number was found just enough to cover the stable phases in the current study. In exploratory searches where the chemistries are unknown, many more structures may need to be looked at. The number of formula units was restricted to being 1 and 2 in the current search. The minimum separation matrix and the cell volumes were determined from a preliminary search which used random specie-pairwise minimum separations of 1-3Å with ~50 structures being sampled on each composition. The details are discussed in section 3.1.1. All the structures generated were constrained to have 2 to 4 symmetry operations.

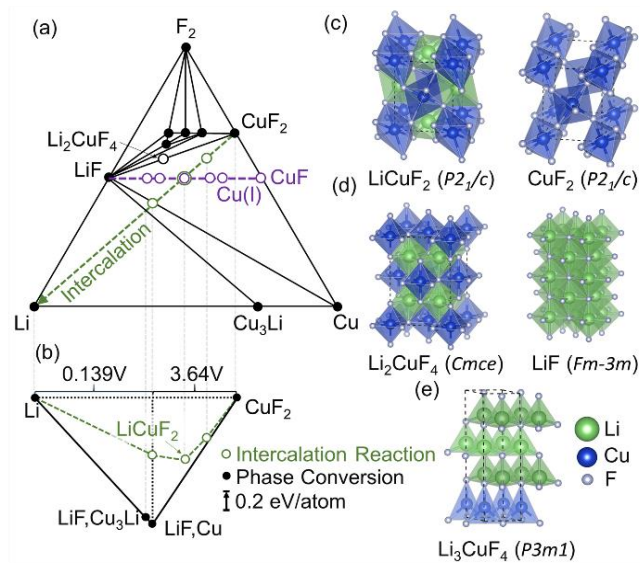

**Fig. 6** (a) Compositional phase diagram for Li-Cu-F. (b) Phase stability along the Li-$CuF_2$ tie line. Crystal structures of (c) $CuF_2$ and the Li-intercalated phase $LiCuF_2$, (d)

LiF and Li$_2$CuF$_4$, and (e) Li$_3$CuF$_4$ with tetrahedrally coordinated Cu (I).

As shown in **Fig. 6(a)**, multiple stable phases were obtained from the search. For the binary compounds, LiF, CuF$_2$ and Cu$_3$Li were found to be stable. Such a result helps determine the terminal redox couple of the CuF$_2$ as a cathode, *i.e.*, the final thermodynamically stable reaction product should be Cu and LiF. The average voltage is calculated to be 3.64 V. For ternary compounds, multiple were found to shape the convex hull, including Li$_2$CuF$_6$, LiCu$_2$F$_6$, and LiCuF$_4$, in close agreement with existing databases.[16] Since none of these stable phases are located within the CuF$_2$-LiF-Cu triangle, CuF$_2$ is expected to adopt a single-step conversion reaction to form LiF and Cu upon lithiation. Interestingly, we also found a number of phases that are less stable (above the convex hull), but are potentially interesting based on kinetics. For example, the phase stability along the Li-CuF$_2$ tie line is shown in **Fig. 6(b)**. The hypothetical Li-intercalated compound LiCuF$_2$ is high in energy (220 meV/atom above hull). Such an instability explains why CuF$_2$ goes through a conversion reaction, rather than intercalation, when lithiated. We also found phases containing F$^-$ anion lattices that are topotactical with those in LiF, as shown in **Fig. 6(c)**. These have recently been proposed as possible buffer phases between the reactant CuF$_2$ and the reaction product LiF.[53] Cu (I) containing fluorides have also been proposed as potential intermediate phases, especially during the charging process. Xiao and co-workers proposed Cu (I) containing phases with tetrahedrally coordinated Cu following electron diffraction analysis.[52] Our search reveals possible phases that are consistent with this. As shown in **Fig. 6(e)**, the Li$_3$CuF$_4$ phase has an energy above hull of 60 meV/atom. Despite its relatively high energy, the formation of this at interfaces, and under electrochemical conditions, is plausible.

To conclude, a multi-composition search can efficiently generate the phase diagram for prediction of phase stabilities and for analysis of the reaction mechanisms of a cathode.

## 2. Exploring new cathodes

Transition metal oxalates are a family of polyanionic compounds that possess a number of advantageous characteristics as cathodes for batteries, but have been overlooked as compared to other polyanionic materials such as the phosphates $PO_4^{3-}$, sulfates $SO_4^{2-}$, and silicates $SiO_4^{4-}$.[54, 55] Interestingly, the $C_2O_4^{2-}$ anion has a similar polarizability to $PO_4^{3-}$. As a result, it should be able to deliver reasonable redox potentials during lithiation, like $LiFePO_4$. Furthermore, oxalates can be synthesized under mild conditions via solution-based methods. In fact, they are known to nucleate into a chain-like, layered, and fully connected frameworks, thanks to complex ligand coordination chemistry. This offers exciting opportunities for oxalates to serve as cathode materials by tuning the structural features through metastability. Despite these advantages, oxalates have been overlooked as cathode materials, in part due to the large molecular weight of the anion, which leads to relatively low energy density.[55] Recently, with the discovery of reversible anion redox as an additional source of capacity, the oxalates are gaining attention from battery researchers. Jiao *et al.* synthesized a lithium-containing iron oxalate with composition $Li_2Fe(C_2O_4)_2$.[56] In this material, Fe has an oxidation state of +2. Interestingly, they found that more than 1 Li/f.u. can be taken out of the structure reversibly. Using spectroscopic evidence, they found that this over-delithiation is enabled by the reversible electron stripping from the $C_2O_4^{2-}$ anion. As compared to the oxides, especially Li-rich ones, the strong covalent O-C bonds in $C_2O_4^{2-}$ prohibits $O_2^{2-}$-dimer formation and the $O_2$ evolution during anion oxidation.

Given the possible polymorphism and interesting chemistry in Li-stuffed transition metal oxalates, we carried out exploratory AIRSS searches on a range of systems of this nature. In particular, we considered $Li_2TM(C_2O_4)_2$, where TM=Fe, Co, Ni, V, and Mn. The aim was to find low energy phases with the potential to be synthesized and display reasonable energy density and rate capability as cathodes for LIBs.

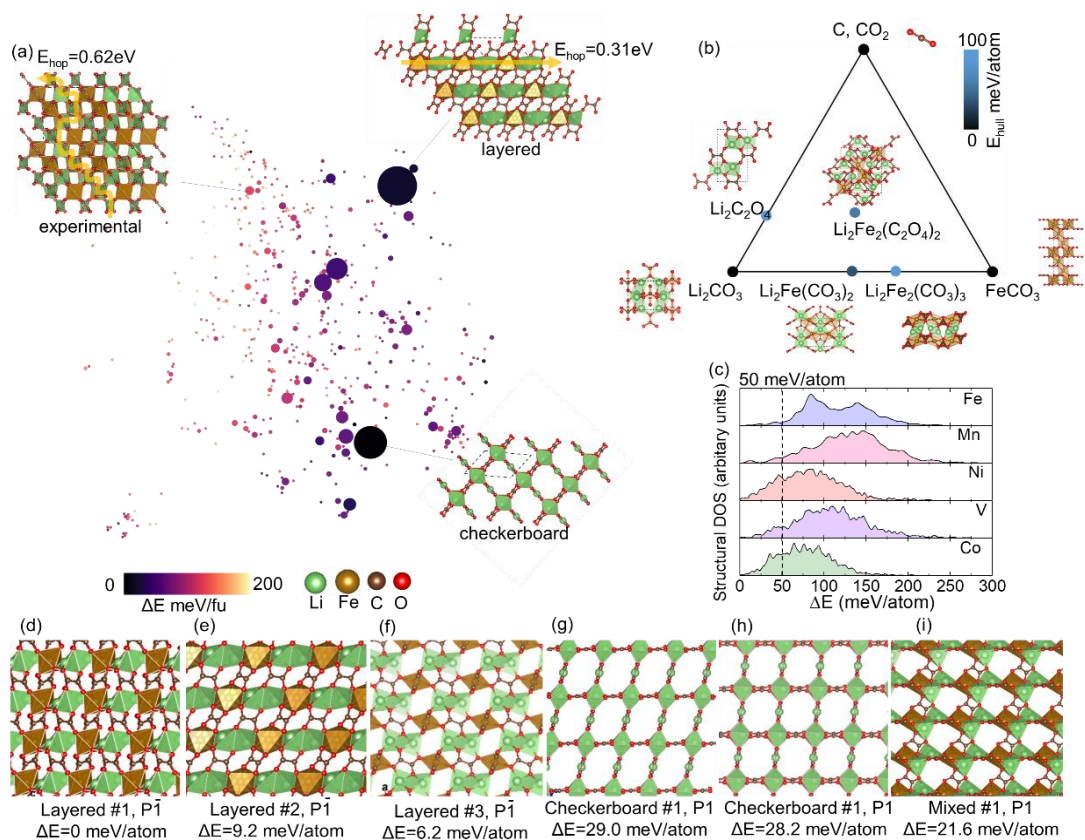

**Fig 7** (a) SHEAP map of AIRSS search results for $Li_2Fe(C_2O_4)_2$. (b) Ternary slice of the Li-Fe-C-O phase diagram containing the decomposition products of $Li_2Fe(C_2O_4)_2$. (c) Structural density of states of the AIRSS search results for $Li_2TM(C_2O_4)_2$ where TM=Fe, Co, Ni, V, and Mn. (d)-(i) Structures of the low energy polymorphs of $Li_2Fe(C_2O_4)_2$.

**Fig. 7(a)** summarises our AIRSS searches for $Li_2Fe(C_2O_4)_2$ in the form of a SHEAP map. During the search, the C and O were treated as $C_2O_4$ units when generating the initial random structures, and the number of formula units was restricted to being 1 to 4. As illustrated in the SHEAP map, among the large number of local energy minima found, only a fraction are found to be low in energy, emphasising the complexity of this PES. Representative low energy structures are labelled in **Fig. 7(a)**. The structures with the highest encounter rates are layered in which the Li$^+$ cations are distributed within the layers. These structures resemble the high-temperature phase of LiCoO$_2$. A number of "checkerboard" structures, composed of one-dimensional structural units of octahedral-coordinated Fe, and planar square-coordinated Li, were also encountered frequently.

**Table 2** lists the energetics of the low energy phases of Li$_2$TM(C$_2$O$_4$)$_2$, whose structures are given in **Figs. 7(d)-7(i)** using Fe compounds as an example. Interestingly, for Fe, Co, and Ni, we found that at least one of the layered polymorphs have an energy lower than the experimentally reported structure at the PBE level. For Fe, this polymorph is characterized by its well-connected Li-network, and is expected to conduct lithium more efficiently than the experimental phase. [57] This is confirmed through calculation of the diffusion barrier of Li in the two structures. In the layered structure, the barrier for Li diffusion ~320 meV, close to some superionic conductors, whereas for the experimentally reported structure, the value is ~620 meV. [56, 58] This indicates that the layered phase has the potential to display high-rate capability as a cathode.

In our search, as well as this layered ground-state structure, a number of metastable structures were found that may be synthesizable through complex nucleation kinetics in solution. Indeed, we calculated that the experimentally reported Li$_2$Fe(C$_2$O$_4$)$_2$ phase is also intrinsically unstable, with an energy above hull of 58 meV/atom. Here, we used such a value as a reference and increased it slightly (by 20 meV/atom) as the threshold for selecting Li$_2$TM(C$_2$O$_4$)$_2$ structures on which to conduct further property calculations. These are the structures before the dashed line in **Fig. 7(c)**. The average discharge voltages of structures below this threshold are listed in **Table 2**. For Fe, all of them are over 4V versus Li/Li$^+$, as expected, since anion redox is involved. We estimated upper bounds for capacities of these materials by checking the structural integrity of the delithiated phases at different states of charge. We found that when the structure is over-delithiated, the $C_2O_4^{2-}$ anion decomposes into two $CO_2$ molecules. Using whether such decomposition happens during structural relaxation as a criterion, the estimated capacity upper bound is listed in **Table 2**. Interestingly, among the TMs we considered, V-oxalates seemed to give the highest capacity without structural collapse, whereas Ni-oxalates easily decomposed when delithiated. The stability of V-oxalates may be due to the multi-valence nature of V. Further, by multiplying the capacity and the voltage, we computed an estimate for the energy density of oxalates. A number of polymorphs (Fe-Layered#1, Mn-Layered#2, Mn-Mixed, Co-Layered#2, and all V-polymorphs) appear to have the potential to reach extraordinary gravimetric energy densities (> 900 Wh kg$^{-1}$), close to or even higher than that of LiNi$_{0.8}$Co$_{0.1}$Mn$_{0.1}$O$_2$.

**Table 2** Properties of Li$_2$TM(C$_2$O$_4$)$_2$ polymorphs within an energy threshold ΔE of 20 meV/atom with respect to the most stable phase found. Here, ΔE is the relative energy compared with the most stable polymorph, E$_g$ is the electronic band gap, and V$_a$ is the average voltage versus Li/Li$^+$. "Exp" and "CB" are abbreviations for "experimental" and "checkerboard", respectively.

| TM | Polymorph | ΔE (meV/atom) | E$_g$ (eV) | V$_a$ (V) | Maximum charge state |
|---|---|---|---|---|---|
| Fe (II) | ***Layered #1*** | ***0*** | ***2.68*** | ***4.18*** | ***Li$_0$Fe(C$_2$O$_4$)$_2$*** |
| | Exp. | 1.4 | 3.03 | 4.24 | Li$_{0.75}$Fe(C$_2$O$_4$)$_2$ |
| | Layered #2 | 6.2 | 2.80 | 4.08 | Li$_{0.5}$Fe(C$_2$O$_4$)$_2$ |
| | Layered #3 | 9.2 | 2.75 | 4.01 | Li$_{0.5}$Fe(C$_2$O$_4$)$_2$ |
| Mn (II) | ***Exp.*** | ***0*** | ***3.86*** | ***4.48*** | ***Li$_{0.75}$Mn(C$_2$O$_4$)$_2$*** |
| | Layered #1 | 12.1 | 3.37 | 4.30 | Li$_{0.75}$Mn(C$_2$O$_4$)$_2$ |
| | Layered #2 | 10.0 | 3.56 | 4.58 | Li$_0$Mn(C$_2$O$_4$)$_2$ |
|  | Layered #3 | 11.3 | 3.49 | 4.35 | $Li_{0.5}Mn(C_2O_4)_2$ |
| Co (II) | **Layered #3** | **0** | **3.96** | **4.70** | **$Li_0Co(C_2O_4)_2$** |
|  | Exp. | 1.4 | 4.21 | 3.74 | $Li_{0.75}Co(C_2O_4)_2$ |
|  | Layered #2 | 7.1 | 4.15 | 4.11 | $Li_0Co(C_2O_4)_2$ |
|  | Layered #1 | 9.3 | 4.18 | 4.20 | $Li_1Co(C_2O_4)_2$ |
|  | CB. #1 | 20 | 4.12 | 4.19 | $Li_{0.5}Co(C_2O_4)_2$ |
| Ni (II) | **Layered #1** | **0** | **4.52** | **4.50** | **$Li_1Ni(C_2O_4)_2$** |
|  | Layered #2 | 1.3 | 4.63 | 4.49 | $Li_1Ni(C_2O_4)_2$ |
|  | Layered #3 | 3.3 | 4.48 | 4.49 | $Li_1Ni(C_2O_4)_2$ |
|  | Exp. | 8.8 | 4.73 | 3.76 | $Li_1Ni(C_2O_4)_2$ |
|  | CB. #2 | 15.6 | 4.47 | 4.47 | $Li_1Ni(C_2O_4)_2$ |
|  | mixed | 19.8 | 4.24 | 4.42 | $Li_1Ni(C_2O_4)_2$ |
| V (II) | **Exp.** | **0** | **2.28** | **4.30** | **$Li_0V(C_2O_4)_2$** |
|  | Layered #2 | 0.8 | 2.13 | 3.85 | $Li_0V(C_2O_4)_2$ |
|  | Layered #3 | 2.7 | 2.05 | 3.58 | $Li_0V(C_2O_4)_2$ |
|  | CB. #1 | 19.5 | 2.01 | 4.40 | $Li_0V(C_2O_4)_2$ |

## IV. CONCLUSIONS

We have explored *ab initio* random structure searching as a tool for battery cathode discovery. We showed, through case studies of known cathode materials of $LiCoO_2$, $LiFePO_4$, and $Li_xCu_yF_z$, that efficient prediction of the low energy structures can be achieved. We demonstrated that using species dependent minimum interatomic separations and cell volumes to steer the search is critical to its efficiency. Further acceleration was achieved by imposing symmetry constraints, and by use of pre-defined chemical subunits for structure generation. Based on these principles, we have carried out explorative searches for $Li_2TM(C_2O_4)_2$ oxalates. A number of low energy polymorphs with layered structures were found. Post-search screening and property calculations identified promising cathode materials characterized by high voltages (> 4V) and low Li diffusion barriers (~ 300 meV), with the potential to deliver not only high energy densities (up to ~900Wh kg$^{-1}$), but also high-rate capability. The current work serves as a basis for AIRSS-based cathode searches, providing a detailed framework for an efficient workflow.

**ACKNOWLEDGEMENTS**

This work was supported by the Faraday Institution grant number FIRG017 and used the MICHAEL computing facilities. C. J. P. acknowledges support from the EPSRC through the UKCP consortium (Grant EP/ P022596/1). D. O. S. acknowledges support from the European Research Council, ERC (Grant 758345). Via membership of the UK's HEC Materials Chemistry Consortium, which is funded by the EPSRC (EP/L000202, EP/R029431), this work used the ARCHER UK National Supercomputing Service (www.archer.ac.uk) and the UK Materials and Molecular Modelling (MMM) Hub (Thomas | EP/P020194 & Young | EP/T022213). B. S. acknowledges EPSRC CDT in Computational Methods for Materials Science for funding under grant number EP/L015552/1.

**DATA AVAILABILITY**

The data that support the findings of this study are available from the corresponding author upon request.